\documentclass[a4paper,11pt]{article}

\pdfoutput=1 

\usepackage{jcappub} 

\usepackage[T1]{fontenc} 
\usepackage{array}
\usepackage{booktabs}
\usepackage{enumitem}
\usepackage[dvipsnames]{xcolor}
\usepackage{tikz}
\usepackage{color}

\title{Forecasts for interacting dark energy with time-dependent momentum exchange}

\author[a]{Nathan Cruickshank,}
\author[a]{Robert Crittenden,}
\author[a]{Kazuya Koyama,}
\author[a]{and Marco Bruni}

\affiliation[a]{Institute of Cosmology and Gravitation, University of Portsmouth,\\
Dennis Sciama Building, Burnaby Road, Portsmouth, PO1 3FX, United Kingdom}

\emailAdd{nathan.cruickshank@port.ac.uk}
\emailAdd{robert.crittenden@port.ac.uk}
\emailAdd{kazuya.koyama@port.ac.uk}
\emailAdd{marco.bruni@port.ac.uk}

\abstract{Models of interacting dark energy and dark matter offer a possible solution to cosmological tensions. In this work, we examine a pure momentum-exchange model with a time-dependent coupling strength $\xi(z)$ that could help to alleviate the $S_8$ tension. We perform Fisher forecasting and MCMC analysis to constrain the coupling strength of this interaction for different redshift bins $0.0<z<2.1$, using the specifications of upcoming DESI-like surveys. For this analysis, we examine both a model with a constant equation of state $w=-0.9$, as well as a thawing dark energy model with an evolving $w(z)$. We show that, for a constant equation of state, $\xi(z)$ can be well constrained in all redshift bins. However, due to a weaker effect at early times, the constraints are significantly reduced at high redshifts in the case of a thawing $w(z)$ model.}

\begin{document}
\maketitle
\flushbottom

\section{Introduction}
\label{sec:intro}

The standard cosmological model $\Lambda$CDM represents our 
best understanding of the history and dynamics of the largest structures around us. This model includes dark energy (DE) in the form of a cosmological constant {$\Lambda$}, which is used to explain measurements of Type Ia supernovae (SNe Ia) \cite{Riess_1998} that indicate that the expansion rate is accelerating; it also includes cold dark matter (CDM), which is necessary to explain measurements of the rotation curves of galaxies and gravitational lensing observations \cite{rotation_curve,Clowe_2004}. Despite the successes of this model, as measurements of large-scale structure (LSS) and the cosmic microwave background (CMB) have improved over time, apparent tensions have started to appear between late-time observations and the predictions of $\Lambda$CDM. 

One prominent discrepancy, known as the Hubble (\textit{$H_0$}) tension, is the $5\sigma$ disagreement between late-time measurements of the Universe's expansion rate and that inferred from CMB and LSS measurements assuming {flat (zero spatial curvature)} $\Lambda$CDM \cite{hu2023hubble,Abdalla:2022yfr,Riess_2016,Riess_2022}. The other key tension, which this work will focus on, relates to measurements of $\sigma_8$ or $S_8$, which quantify the degree of clustering in the Universe. Similarly to the \textit{$H_0$} tension, improvements in the precision of measurements of the CMB and later measurements of LSS have highlighted tensions between the data sets. A $2-3\sigma$ tension has  been seen between the clustering amplitude predicted from Planck measurements and more direct late-time probes, such as galaxy lensing \cite{Planck_2018,Hildebrandt_2020,Asgari_2021,Di_Valentino_2021,Abbott_2022}. These late-time measurements show a preference for a lower value of $S_8$ and less structure growth than the CMB would imply for a $\Lambda$CDM model.

These tensions give an indication that $\Lambda$CDM may be wrong or incomplete; as a result, various solutions have been proposed. Models of interacting dark energy and dark matter provide one potential solution \cite{Bolotin_2015,Wang_2016,Abdalla:2022yfr,Tamanini_2015,Pourtsidou_2016,Di_Valentino_2017,Pan_2019,Pan_2020_1,Pan_2020_2,Amendola_2020,Di_Valentino_2020_1,Di_Valentino_2020_2,Khoury:2025txd}. These propose that, unlike in $\Lambda$CDM, the components of the dark sector can interact and affect one another in ways other than gravitational. This is despite neither appearing to interact strongly with ordinary matter. 

In models involving the exchange of energy or energy-momentum, both the background and perturbation dynamics are affected. Such interactions can lead to shifting acoustic peak positions as well as Integrated Sachs-Wolfe (ISW) effect signatures in the CMB power spectrum \cite{Xia_2009,Amendola_2012,G_mez_Valent_2020}. In this work, we do not explore such models; however, recent work has utilised data, such as CMB measurements and the latest data from the Dark Energy Spectroscopic Instrument's (DESI) Baryon Acoustic Oscillations (BAO) observations, to constrain these interactions \cite{Sabogal:2024yha,Li:2024qso,Wang:2024hks,Ghedini:2024mdu,Pan:2025qwy}. This data has also been used to explore how such interactions could be used to resolve the $H_0$ tension. It has been found that the simplest models can solve the $H_0$ tension but result in undesirable effects, such as a worsened $S_8$ tension \cite{DiValentino:2020zio,Gariazzo:2021qtg,Abdalla:2022yfr,Hoerning:2023hks,Giare:2024smz,Giare:2024ytc}. More complex models, such as sign-switching interactions, are required to alleviate both tensions simultaneously \cite{Silva:2025hxw,Alexander:2022own,Shah:2024rme,Sabogal:2025mkp}.

Alternatively, in models where only momentum is exchanged, the background evolution is not directly affected, and for this reason, such an interaction cannot solve the $H_0$ tension on its own. However, these models can affect the evolution of perturbations, allowing them to potentially resolve the $S_8$ tension while also fitting CMB data. Crucially, pure momentum exchange models require that DE not be modelled as a vacuum in order for momentum transfer to occur \cite{Wands:2012vg}. For this reason, the background evolution must necessarily be different from $\Lambda$CDM. Dynamical DE has been proposed as one potential solution to the $H_0$ tension \cite{DiValentino:2021izs}. This means it may be possible to tune the complete model to alleviate both the $H_0$ and $S_8$ tensions simultaneously. Studies of momentum exchange models have taken two approaches, either more fundamental models often derived from a Lagrangian, or more phenomenological approaches. An example of the former is coupled quintessence models with pure momentum exchange \cite{Pourtsidou_2013,Skordis_2015,Pourtsidou_2016,palma2023cosmological}. Constraining such models, while very useful, can also be hard to generalise.

In this paper, we consider a more phenomenological interaction model that captures the essential physics, one that has elastic scattering with pure momentum exchange. We perform a Fisher forecast in order to test how well a DESI-like, Stage IV dark energy survey will be able to constrain the strength of the interaction. We emphasise that we do not use any actual DESI observations in this analysis, but forecast what such a survey may eventually be capable of.   This analysis is conducted using a constant dark energy equation of state $w\equiv p/\rho$, where $p$ and $\rho$ are the dark energy pressure and energy density respectively, as well as an evolving $w(z)$ parametrisation.

The outline of this paper is as follows: in Sec. \ref{sec2}, we describe the interaction model as well as our chosen $w(z)$ parametrisation. In Sec. \ref{sec3}, we describe the methods used in our analysis as well as the parameter values of our fiducial cosmology. In Sec. \ref{sec4}, we present the results of our Fisher forecasting analysis when constraining the strength of the interaction. We conclude in Sec. \ref{sec5}.

\section{Theoretical models}
\label{sec2}

\subsection{Elastic scattering momentum exchange}

The focus of this work will be the elastic scattering model proposed in \cite{Simpson_2010}. This interaction model is an idealised case of the more general class of momentum exchange models; however, it is shown in \cite{palma2023cosmological} using N-body simulations that, in the limit of weak coupling, general momentum exchange models are consistent with the dark scattering case. Developed using the $w$CDM extension to $\Lambda$CDM, this model involves purely momentum transfer, inspired by Thompson scattering. Due to CDM's non-relativistic velocities and the low density of DE, the interaction can be expected to consist of slow, low energy impacts that maintain elasticity. When we set $c=1$, the interaction strength of this model is determined by the value of the parameter $\xi=\sigma_\mathrm{D}/m_\mathrm{cdm}$ [b/GeV]. This parameter represents the ratio of the DE-CDM interaction cross-section, $\sigma_\mathrm{D}$, to the mass of a CDM particle, $m_\mathrm{cdm}$. The model has been shown to affect late-time structure formation through the use of Einstein-Boltzmann solvers, N-body simulations and emulators \cite{Simpson_2010,Baldi_2016,Kumar:2017bpv,Bose_2018,Asghari_2020,Beltr_n_Jim_nez_2021,Carrilho_2021,Linton_2022,Poulin_2023,Carrilho_2023,laguë_2024,Carrion:2024jur}. 

The effect of such a momentum exchange on the growth of structure depends on the DE background evolution and on the scales considered. Dark scattering features an explicit dependence on the value of $w$. This is an intrinsic feature of a model where CDM particles move through a fluid with equation of state $w$ and are elastically scattered. When $\xi > 0$ and $w > -1$ ($w < -1$), there is an observed suppression (enhancement) in the structure growth rate on linear scales as a result of the introduced friction between DE and CDM. 
This description holds true on linear scales; however, it has been shown by numerical simulations that the inverse effect is observed on non-linear scales. At these smaller scales, when $\xi > 0$ and $w > -1$ ($w < -1$), the growth of structure is strongly enhanced (suppressed) by the momentum exchange \cite{Baldi_2015}. This results from the friction term increasing the efficiency of gravitational collapse by lowering the kinetic energy of the bound structures. The observed suppression on linear scales for  $w > -1$ demonstrates how the interaction can help to alleviate the $S_8$ tension, where $S_8\equiv\sigma_8\sqrt{\Omega_\mathrm{m}/0.3}$. Here, $\sigma_8$ describes the amplitude of matter perturbations in an $8h^{-1}\mathrm{Mpc}$ radius sphere and $\Omega_\mathrm{m}$ is the matter density parameter. An alternative parametrisation, introduced in \cite{Asghari:2019qld} and subsequently explored in \cite{Beltr_n_Jim_nez_2021,Figueruelo:2021elm,BeltranJimenez:2022irm,Tsedrik:2022cri,BeltranJimenez:2024lml}, has also been shown to suppress structure growth and alleviate the $S_8$ tension.

Previous literature has evaluated this model using Markov Chain Monte Carlo (MCMC) analysis and emulators to constrain the model's parameters with current CMB and LSS data sets. One parametrisation that has been examined is $A\equiv\xi\left(1+w\right)$, which is directly related to the size of the damping term and encodes the strength of the interaction on observables \cite{Linton_2022,Carrilho_2021,carrilho_20232,carrion_2024,tsedrik_2025}. This parametrisation helps avoid a very large value of $\xi$ when analysing a cosmology with a value of $w$ close to $-1$. Other work has focused on forecasting the possible constraints on interacting model parameters for upcoming spectroscopic and photometric redshift surveys \cite{Asghari_2020}.  

Although research has been done into the effects of time-dependent couplings in the context of dark energy as a scalar field and with a varying equation of state, this is the first in the context of the phenomenological elastic scattering model \cite{Baldi_2010,Bandyopadhyay_2021,Yang_2018}. In this work, we use redshift binning to allow for the value of $\xi$ to change with time. Rather than using current data sets, we perform a Fisher forecasting analysis to find the constraining power of upcoming DESI-like surveys on these interaction strength parameters.

\subsection{Modified equations}

In the conformal Newtonian gauge and in the absence of DE-CDM coupling, when the fluid sound speed $c_s^2=1$, the continuity and Euler equations for the fluids take the form
\begin{equation}
\label{eq:1}
    \delta^\prime_\mathrm{cdm}=-\theta_\mathrm{cdm}+3\Phi^\prime\:,
\end{equation}
\begin{equation}
\label{eq:2}
    \theta^\prime_\mathrm{cdm}=-H\theta_\mathrm{cdm}+k^2\Psi\:,
\end{equation}
\begin{equation}
\label{eq:de1}
    \delta'_\mathrm{de} = - \left[ (1 + w) + 9 \frac{H^2}{k^2} \left( 1 - w^2 \right) \right] \theta_\mathrm{de} + 3(1 + w) \Phi' - 3H(1 - w) \delta_\mathrm{de}\:,
\end{equation}
\begin{equation}
\label{eq:de2}
    \theta^\prime_\mathrm{de}=2H\theta_\mathrm{de}+\frac{\delta_\mathrm{de}}{\left(1+w\right)}k^2+k^2\Psi\:,
\end{equation}
where $k$ is the wave number and $H$ is the Hubble parameter. $\delta_\mathrm{cdm}$ and $\delta_\mathrm{de}$ are the CDM and DE density contrasts, $\theta_\mathrm{cdm}$ and $\theta_\mathrm{de}$ are the CDM and DE velocity divergences, and $\Phi$ and $\Psi$ are the spatial curvature potential and Newtonian gravitational potential respectively. 
 
The inclusion of the DE-CDM momentum interaction results in an additional term appearing in the DE Euler equation
\begin{equation}
\label{eq:3}
   \theta^\prime_\mathrm{de}=2H\theta_\mathrm{de}+\frac{\delta_\mathrm{de}}{\left(1+w\right)}k^2+a\rho_\mathrm{cdm}\left(\theta_\mathrm{cdm}-\theta_\mathrm{de}\right)\xi(z)+k^2\Psi\:,
\end{equation}
where $a$ is the scale factor and $\rho_\mathrm{cdm}$ is the density of CDM. Conservation of momentum leads to a similar term arising in the CDM Euler equation with a dependence on $1+w$:
\begin{equation}
\label{eq:4}
    \theta^\prime_\mathrm{cdm}=-H\theta_\mathrm{cdm}+a\left(1+w\right)\rho_\mathrm{de}\left(\theta_\mathrm{de}-\theta_\mathrm{cdm}\right)\xi(z)+k^2\Psi\:.
\end{equation}
It can be seen from Equations \ref{eq:1} and \ref{eq:4} that the growth of CDM perturbations can be affected by the inclusion of a non-zero $\xi(z)$. To allow for the interaction to take place, $w(z)$ must also have a value different from $-1$. Physically, as the ratio of a cross-section and a mass cannot be negative, the coupling has to be positive ($\xi(z) \ge 0$) to ensure the stability of perturbations. In this work, we focus on non-phantom DE ($w(z) \ge -1$) and, therefore, impose the condition $\xi(1+w) > 0$. A negative value for this term is possible when allowing $\xi > 0$ and $w < -1$; however, the combination $\xi < 0$ and $w > -1$ would result in runaway growth for the perturbations. This focus on $w(z) \ge -1$ also helps to avoid the well-known phantom crossing instabilities in fluid-based DE models \cite{Hu:2004kh}. In this stable regime, the coupling has the effect of adding a friction term that suppresses the growth of structure.

These effects can be observed through redshift space distortion (RSD) measurements of $f\sigma_8$, where $f(a)$ is the logarithmic growth rate $f(a)\equiv\frac{\mathrm{d}\ln{D}}{\mathrm{d}\ln{a}}$ and $D(a)=\frac{\delta(a)}{\delta(a=1)}$. In order to study the effects of the interaction model, we have implemented the modified Euler equations into the Einstein-Boltzmann solver code \texttt{CLASS} \cite{Blas_2011}. The code enables us to simulate the evolution of $f\sigma_8$ for a time-dependent coupling $\xi(z).$

\begin{table}[tbp]
\centering
\begin{tabular}{|lr|c|}
\hline
Parameter&Redshift Range\\
\hline
$\xi_1$ & $0.0<z<0.4$\\
$\xi_2$ & $0.4<z<1.1$\\
$\xi_3$ & $1.1<z<1.6$\\
$\xi_4$ & $1.6<z<2.1$\\
$\xi_{\mathrm{high}}$ & $2.1<z<10$\\
\hline
$\xi_{\mathrm{low}}$ & $0.0<z<2.1$\\
$\xi_{\mathrm{high}}$ & $2.1<z<10$\\
\hline
\end{tabular}
\caption{\label{tab:i} We consider a time-dependent coupling parameter by binning in redshift; for simplicity, we follow the DESI binning, with an additional high redshift bin to take into account uncertainty in the clustering amplitude from the early universe.  For comparison, we also consider a simpler two parameter case.}
\end{table}

\begin{figure}[tbp]
     \centering 
     \includegraphics[width=0.49\textwidth]{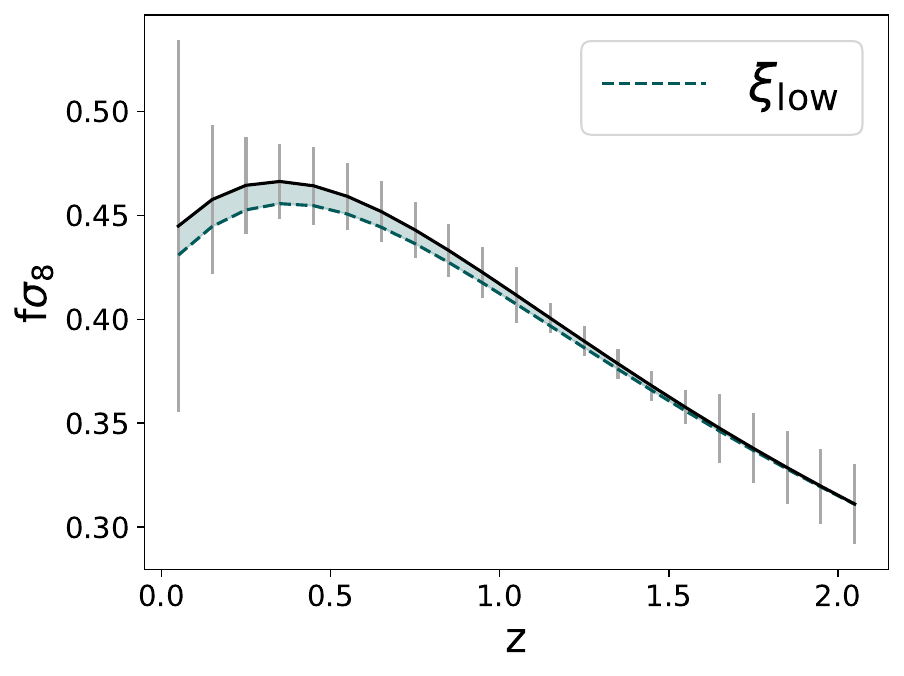}
     \includegraphics[width=0.49\textwidth]{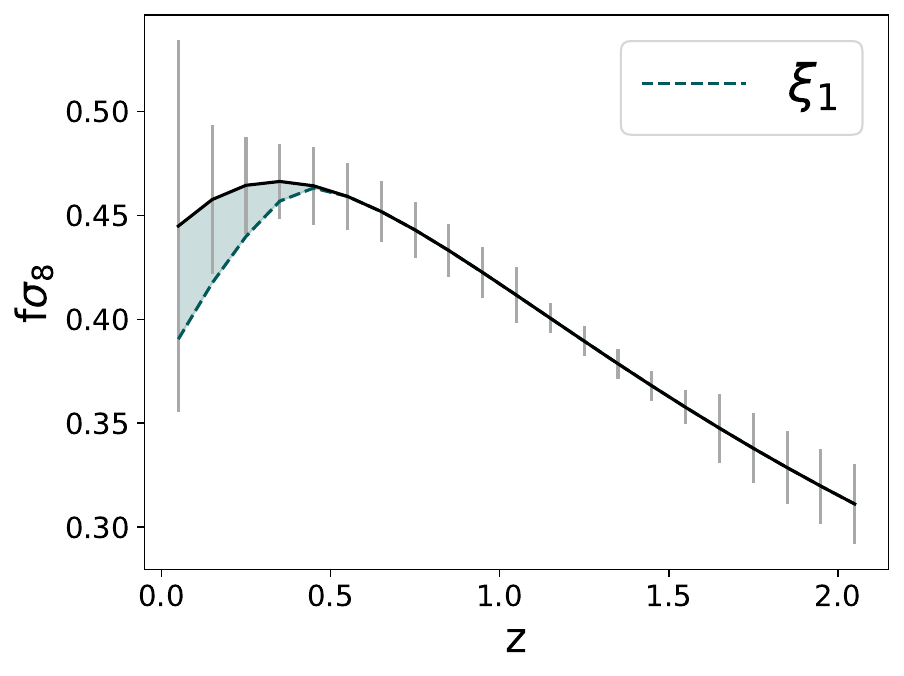}
     \includegraphics[width=0.49\textwidth]{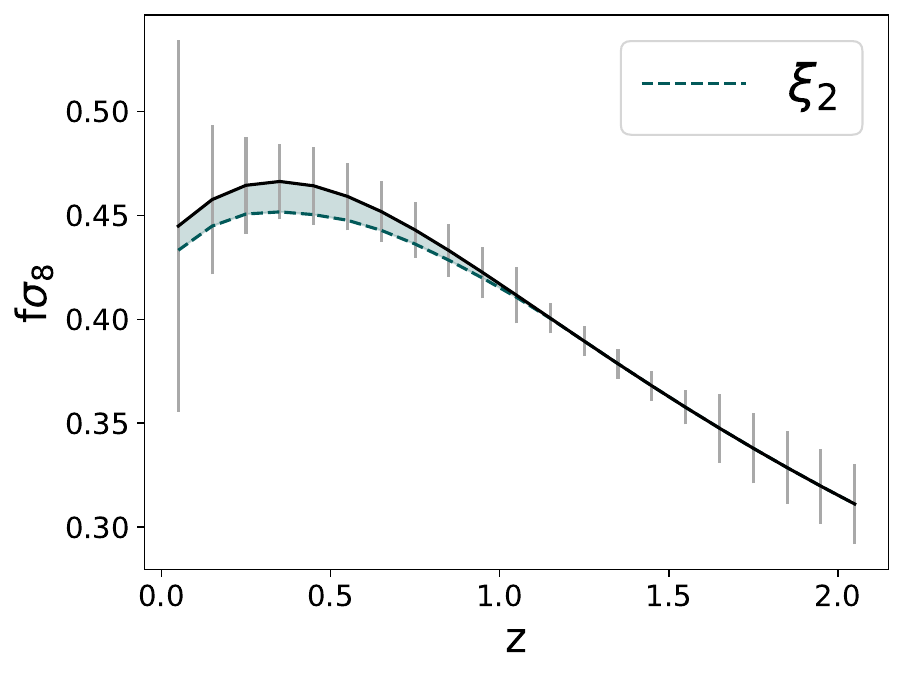}
     \includegraphics[width=0.49\textwidth]{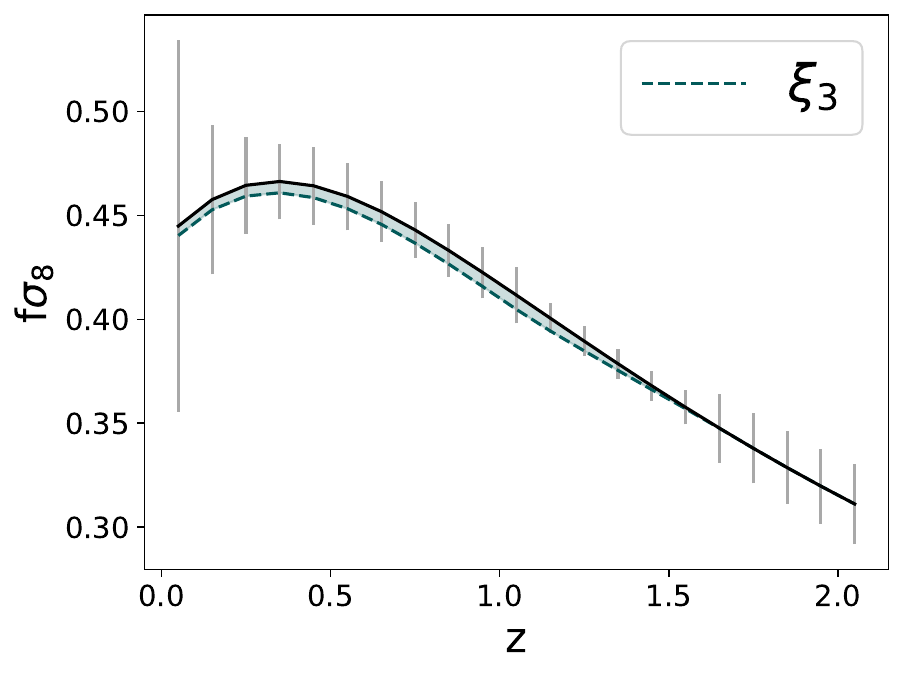}
     \includegraphics[width=0.49\textwidth]{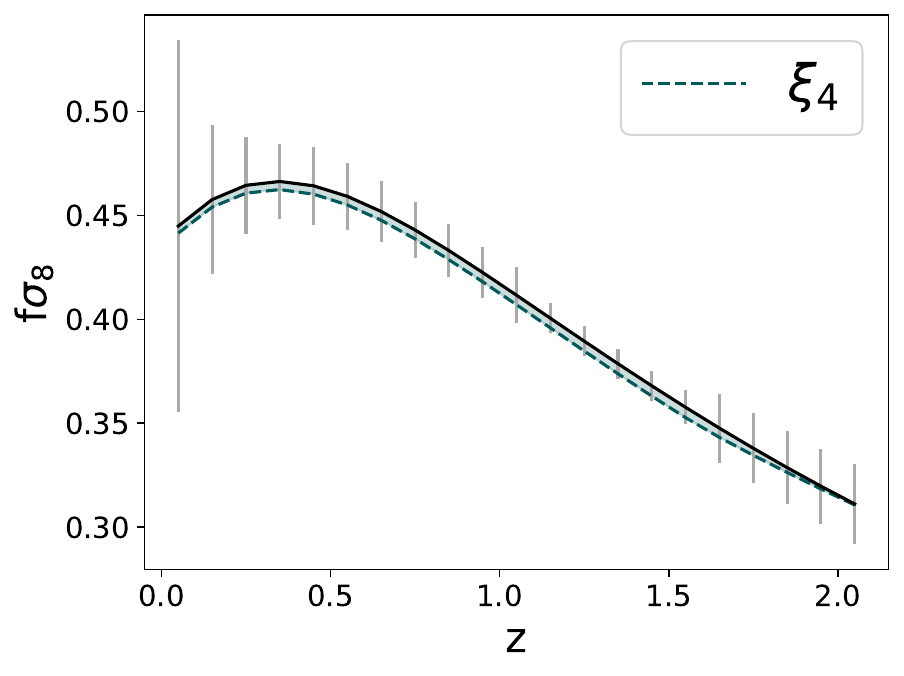}
     \includegraphics[width=0.49\textwidth]{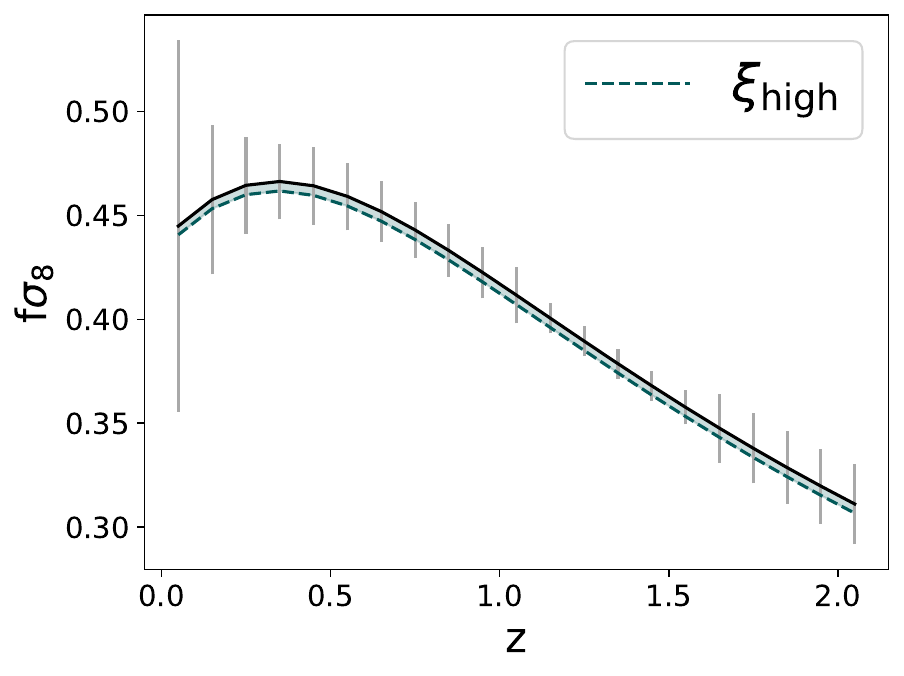}
     \hfill
     \caption{\label{fig:fsigma8} The impact on $f\sigma_{8}$ from changing the coupling in each redshift bin. The error bars are derived from a DESI-like survey, and we show the impact when using the calculated $\xi_{\mathrm{low}}$ and $\xi_i$ $3\sigma$ error values derived below. (See Table \ref{tab:iii}.) These figures used a constant equation of state $w = -0.9$.}
\end{figure}

For our analysis, we consider the effects of binning the coupling $\xi$ in redshift. We first consider a coupling constant at low redshift: $\xi_{\mathrm{low}}$, for $0 < z < 2.1$ and $\xi_{\mathrm{high}}$, for $2.1 < z < 10$. We then further divide $\xi_{\mathrm{low}}$ into four different redshift bins, 
$\xi_{1-4}$; these ranges are given in Table \ref{tab:i}. Our method of redshift binning uses a series of simple functions with steep gradients to smoothly transition between the chosen $\xi$ value of each bin. This method was chosen in order to avoid any numerical instabilities that may arise from an instant transition.

The effect that the coupling in each redshift bin has on the evolution of $f\sigma_8$ can be seen in Figure \ref{fig:fsigma8}. We show the couplings that could be detected at the $3\sigma$ level, which we derive below. As can be seen, the coupling only impacts the growth rate within the bin and at lower redshifts.

\subsection{\texorpdfstring{$w(z)$ parametrisation}{w(z) parametrisation}}

As the interaction is dependent on the dark energy equation of state, we explore the effects of different $w(z)$ models on the DE-CDM interaction constraints. We start with a constant $w_0=-0.9$ model, which though not strongly physically motivated, does allow us to isolate the effects of the interactions. 

We also explore more dynamically evolving dark energy models. Recent BAO data from DESI DR2 measurements, combined with supernovae (SNe) and other data sets, has shown evidence for a time-evolving equation of state \cite{DESI:2025zgx}. This has usually been parameterised using the Chevallier-Polarski-Linder (CPL) parameters $w_0$ and $w_a$, where $w(a) = w_0 + w_a (1-a)$ \cite{Chevallier:2000qy,Linder:2002et}. The values depend on the SNe data set used; for DESI+CMB+DESY5 SNe Ia data, the best fit is $w_0=-0.752\pm0.057$ and $w_a=-0.86^{+0.23}_{-0.20}$. Such a parameterisation is arguably unphysical in a single scalar field model with the standard kinetic term, as the equation of state is phantom $w < -1$ at higher redshifts, where the density of dark energy would actually increase as the Universe expands. For the interacting models considered here, it would also have an effectively negative friction, causing the structure formation to happen faster rather than suppressing it.  

\begin{figure}[tbp]
     \centering 
     \includegraphics[width=0.8\textwidth]{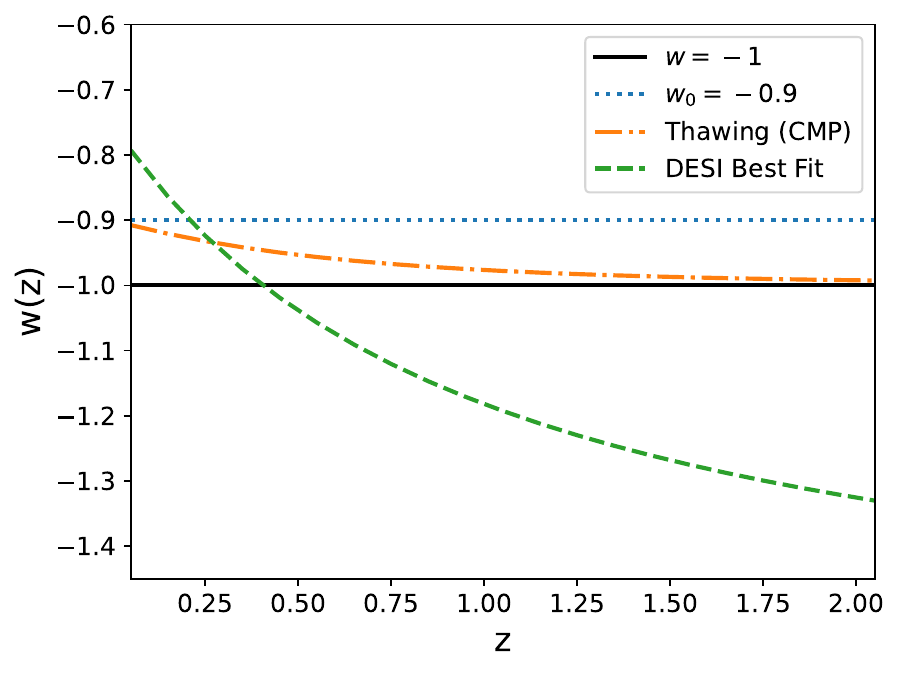}
     \hfill
     \caption{\label{fig:wz} The dark energy equation of state, $w(z)$. We examine the interaction constraints for constant $w_0 = -0.9$ (blue dotted) and the CMP parametrisation motivated by thawing quintessence models (orange, dot-dashed). For comparison, we also plot the best fit $w_0$-$w_a$ model found by DESI using DESI+CMB+DESY5 SNe Ia data \cite{DESI:2025zgx} ($w_0=-0.752\pm0.057$ and $w_a=-0.86^{+0.23}_{-0.20}$). This exhibits phantom behaviour at high redshift.}
\end{figure}

In order to avoid such effects, we consider $w(a)$ parametrisations that are motivated by thawing quintessence models, where the equation of state is always $w > -1$. In the early Universe, the quintessence field is effectively frozen on its potential, acting as a cosmological constant. Only at late times does the field begin to become dynamical, its equation of state slowly increasing but still remaining negative. 
To more accurately model a thawing quintessence field, we follow the parametrisation outlined in Crittenden, Majerotto and Piazza (CMP) \cite{Crittenden_2007}. It is a two-parameter model that exactly reproduces the thawing behaviour in the limit where the equation of state is close to $w=-1$. In it, the equation of state is related to a new function $\kappa(a)$ by 
\begin{equation}
\label{eq:7}
   1 + w(a) \equiv \frac{2}{3}\kappa^2(a)\Omega_\Lambda(a)\:,
\end{equation}
where $\Omega_\Lambda(a) \equiv \frac{\Omega_{\Lambda,0}}{\Omega_{\Lambda,0} + (1-\Omega_{\Lambda,0}) a^{-3}}$
would be the density parameter for a cosmological constant and $\kappa(a)$ is a two-parameter function defined by  
\begin{equation}
\label{eq:6}
    \kappa(a)=\kappa_0\left(1-\Omega_{\Lambda,0}+\Omega_{\Lambda,0}a^3\right)^{-2\kappa_1/3}\:.
\end{equation}
As shown in Figure \ref{fig:wz}, this parametrisation avoids the problem of crossing into the phantom regime ($w<-1$) that emerges from the $w_0$ and $w_a$ parametrisation. The two parameters $\kappa_0$ and $\kappa_1$ can be chosen to match a choice of $w_0$ and $w_a$ at the present. In our \texttt{CLASS} implementation, we also make use of the derivative and integral of $w(a)$
\begin{equation}
\label{eq:8}
    \frac{dw}{da}=\frac{2}{3a}\kappa^2(a)\Omega_\Lambda(a)\left((4\kappa_1+3)\Omega_M(a)-4\kappa_1\right)\:,
\end{equation}
where $\Omega_M(a) = 1 - \Omega_\Lambda(a)$ and 
\begin{equation}
\label{eq:9}
    I(a)= 3 \int_a^1 \frac{ da'}{a'} [1+w(a)] = \frac{\left[\kappa^2-\kappa_0^2\right]}{2\kappa_1}\:.
\end{equation} 
The latter relates directly to the dark energy density, $\rho_{\mathrm{DE}} (a) \propto \exp[I(a)].$

When choosing the values of $\kappa_0$ and $\kappa_1$, we use the thawing equation of state approximation $w_a\approx-1.58\left(1+w_0\right)$ discussed in \cite{DESI:2024kob} and \cite{DESI:2025fii}. With this constraint, they fit to the DESI, CMB and DESY5 SNe data sets, and find that a thawing model with $w_0=-0.9$ fits the data better than the $\Lambda$CDM model.  By matching these values at low redshift, we ensure that the parameterisations match at $z=0$, but the CMP parametrisation never crosses $w=-1$.  However, we specify the models by their equation of state today $w(z=0)$.    
We take this as our fiducial evolving $w(z)$ model;   
the models we consider are plotted in Figure \ref{fig:wz}.  

\section{Forecast methodology}
\label{sec3}

In this section, we allow for the interaction strength to vary across five redshift bins. We examine how well the data from upcoming surveys might be able to constrain the corresponding {$\xi_i=\xi_{1,2,3,4,\mathrm{high}}$ parameters}. We have developed our Fisher forecasting Python code, based on the published pre-DESI forecasts described in \cite{desicollaboration2016desi}.

\subsection{DESI tracers and assumptions}

The size of each $\xi_i$ redshift bin was chosen to match the range of redshifts covered by each DESI-like survey tracer as well an additional bin covering a range up to redshift $10$ \cite{DESI2023}. Each tracer is assumed to have a clustering bias, which describes the relationship between the observed and underlying dark matter density fields. We assume that this evolves as $b_i(z) = b_i/D(z)$ for each different tracer type. The tracers that we are interested in forecasting for include:
\begin{description}[leftmargin=4em,style=nextline]
\item[Bright Galaxy Survey (BGS):] This tracer is a high density sample of the brightest galaxies that exist at low redshift, with a wide range of galaxy properties. 
We treat this as the primary tracer for the redshift range $0.0<z<0.4$. We assume that the bias evolves as $b_{\mathrm{BGS}}(z) = 1.34/D(z)$.
\item[Luminous Red Galaxies (LRGs):] This tracer consists of luminous, massive galaxies
that have largely ceased star formation. We treat this as the primary tracer for the redshift range $0.4<z<1.1$. We assume that the bias evolves as $b_{\mathrm{LRG}}(z) = 1.7/D(z)$.
\item[Emission-Line Galaxies (ELGs):] This tracer consists of galaxies with a high rate of star formation. We treat this as the primary tracer for the redshift range $1.1<z<1.6$. We assume that the bias evolves as $b_{\mathrm{ELG}}(z) = 0.84/D(z)$.
\item[Quasars (QSOs):] This tracer consists of galaxies with quasars, powered by gravitational accretion onto supermassive black holes. We treat this as the primary tracer for the redshift range $1.6<z<2.1$. We assume that the bias evolves as $b_{\mathrm{QSO}}(z) = 1.2/D(z)$.
\end{description}

\subsection{Fisher forecasts}

For $N$ galaxy tracers, the Fisher information matrix can be calculated using
\begin{equation}
\label{eq:10}
    F_{ij} = \sum_{XY}\int \frac{V_0\,d^3k}{(2\pi)^3}
  \ \left( \frac{\partial P_X}{\partial p_i} \right)
  C^{-1}_{XY}
  \left( \frac{\partial P_Y}{\partial p_j} \right)
  \quad\:,
\end{equation}
where $V_0$ is the geometric volume of the survey, $p_i$ is a set of parameters, $C$ is the covariance matrix, X and Y denote a pair of tracer indices and $P$ is the measured power spectrum \cite{White_2009}.

\subsubsection{Volume and mean number density}

We calculate the volume of each redshift increment using
\begin{equation}
\label{eq:13}
	V_i=\frac{4\pi}{3}\,f_\mathrm{sky}\,\big(d_c^3(z_\mathrm{max})-d_c^3(z_\mathrm{min})\big) \:,
\end{equation}
where $f_{sky}$ is the fraction of the sky covered by the survey, which has an area of $14,000$ $\mathrm{deg}^2$, and $d_c$, which is defined as
\begin{equation}
\label{eq:14}
	d_c(z)=\int^z_0 \frac{c}{H(z)}\,\mathrm{d}z \:,
\end{equation}
is the comoving distance to redshift $z$. We also calculate the mean number density $\bar{n}$ of each redshift bin using
\begin{equation}
\label{eq:15}
	\bar{n}_i=\frac{4\pi}{V_i}\,f_\mathrm{sky}\,\int^{z_\mathrm{max}}_{z_\mathrm{min}} \mathrm{d}z \frac{\mathrm{d}N}{\mathrm{d}z}(z) \:,
\end{equation}
where $\frac{\mathrm{d}N}{\mathrm{d}z}(z)$ is the surface number density of the survey.

We use the surface density values given in \cite{DESI2023} for our forecasting calculations. Additionally, for the fiducial cosmology of our analysis, we use the parameter values given in Table \ref{tab:ii}, following the Planck 2018 TT,TE,EE+lowE+lensing results \cite{Planck_2018}. We choose to fix these values as they are reasonably well constrained.

\begin{table}[tbp]
\centering
\begin{tabular}{|lr|c|}
\hline
Parameter&Value\\
\hline
$\omega_b$ & 0.02237\\
$\omega_c$ & 0.1200\\
$\theta_s$ & 1.04110\\
$\log_{10}A_s$ & 3.044\\
$n_s$ & 0.9649\\
$\tau$ & 0.0544\\
\hline
\end{tabular}
\caption{\label{tab:ii} The parameter values for our fiducial cosmology, following the Planck 2018 TT,TE,EE+lowE+lensing results \cite{Planck_2018}.}
\end{table}

\subsubsection{Power spectrum and BAO reconstruction effects}

When using RSD information, the power spectrum of each tracer is assumed to be 
\begin{equation}
\label{eq:11}
    P\left(k,\mu,z\right) = \left(b(z)\sigma_8(z)+f(z)\sigma_8(z)\mu^2\right)^2\frac{P_\mathrm{mass}\left(k,z\right)D_{NL}\left(k,\mu,z\right)}{\sigma_8^2(z)}\:,
\end{equation}
where $\mu$ is the cosine of the angle between the wave-vector $k$ and the line-of-sight direction. The matter power spectrum $P_\mathrm{mass}\left(k,z\right)$ was computed using \texttt{CLASS} and then factored into smooth and BAO components. Following \cite{DESI2023}, this was done to allow for Alcock-Paczyński (AP) projection effects and the degradation of the BAO reconstruction due to shot noise to be taken into account. We compute $f\sigma_8(z)$ as a scale-independent quantity using $\frac{\mathrm{d}\sigma_8}{\mathrm{d}\ln{a}}$. However, we have also explored calculating $f\sigma_8(z)$ as a scale-dependent quantity at a single $k$ value and found comparable results for the models we consider.

AP projection effects describe distortions in the observed clustering of the BAO as a result of using the incorrect cosmology for redshift and angle measurements \cite{Alcock:1979mp}. Distortions in the radial direction depend on $1/H(z)$ and distortions in the angular direction depend on the angular diameter distance $D_\mathrm{A}(z)$. When using a fiducial cosmology to convert redshifts to distances, the parameters $\alpha_\perp(z)=D_A(z)/D_{A,\mathrm{ref}}(z)$ and $\alpha_\parallel(z)=H_{\mathrm{ref}}(z)/H(z)$ can be used to describe the effect. In our calculations, the effect is applied only to the BAO component of the power spectrum. We follow \cite{Ballinger:1996cd} and use 
\begin{equation}
\label{eq:k}
    k\left(k_\mathrm{fid},\mu_\mathrm{fid}\right) = \frac{k_\mathrm{fid}}{\alpha_\perp}\left[1+\mu^2_\mathrm{fid}\left(\frac{\alpha^2_\perp}{\alpha^2_\parallel}-1\right)\right]^{1/2}\:,
\end{equation}
to translate fiducial $k$ and $\mu$ values into real $k$ values.

BAO reconstruction degradation occurs when taking into account BAO uncertainties from non-linear growth. This is calculated by introducing a damping strength factor to the power spectrum as well as a reconstruction factor that is determined by the tracer’s shot noise. The damping factor includes the Lagrangian displacement distances $\Sigma_\perp=9.4\left(\sigma_8(z)/0.9\right)h^{-1}$Mpc and $\Sigma_\parallel=\Sigma_\perp\left(1+f(z)\right)$, which are multiplied by a factor $\in\left[0.5,1\right]$ to account for the degradation of the reconstruction due to shot noise. This effect is considered when modelling RSD distortions using the linear Kaiser model \cite{10.1093/mnras/227.1.1}.  Following \cite{Seo:2007ns}, the damping factor $D_{NL}(k,\mu,z)$ is given by
\begin{equation}
\label{eq:12}
    D_{NL}\left(k,\mu,z\right) = \mathrm{exp}\left[-k^2\left(\frac{\left(1-\mu^2\right)\Sigma_\perp^2}{2}+\frac{\mu^2\Sigma_\parallel^2}{2}\right)\right]\:.
\end{equation}

\subsection{Deriving coupling constraints}

We use Equation \ref{eq:10}, along with the pre-DESI survey tracer assumptions, to calculate the errors on $f\sigma_8(z)$, $b_i\sigma_8(z)$, $\alpha_\perp(z)$ and $\alpha_\parallel(z)$ for redshifts $0.0<z<2.1$ in increments of $0.1$. We are then able to use these results to derive constraints on the coupled model parameters, including the errors on the $\xi_i$ parameters and $w_0$, in the constant and evolving $w$ cases. 

We consider a coordinate transform, where the new parameters $p_i({\bf \theta})$ are defined in terms of the original parameters $\theta_\alpha$.  
We use a Jacobian transform to create a new Fisher matrix, $S_{ij}$ from the original $F_{\alpha\beta}$:
\begin{equation}
\label{eq:16}
    S_{ij}=\sum_{\alpha\beta}{\frac{\partial\theta_{\alpha}}{\partial p_{i}}F_{\alpha\beta}\frac{\partial\theta_{\beta}}{\partial p_{j}}}\:.
\end{equation}  
When transforming the errors on the parameters, the $\alpha_\perp(z)$ and $\alpha_\parallel(z)$ constraints transfer to the background and provide constraints on $w_0$.  The $f\sigma_8(z)$ errors largely map to constraints on the $\xi_i$ parameters. 

We also use MCMC sampling as an alternative method to derive more accurate parameter constraints. Fisher forecasting assumes a Gaussian likelihood, whereas MCMC samples the full posterior distribution, even if it is non-Gaussian. For this, we make use of the Monte Carlo code \texttt{MontePython}, as it is able to easily interface with \texttt{CLASS} \cite{Brinckmann:2018cvx,Audren:2012wb}. For the MCMC analyses, we chose to use flat priors on the couplings and enforce positivity $\xi \ge 0$ on each $\xi_i$ parameter;  
similarly, for the equation of state, we require $w>-1$.

\section{Results and discussion}
\label{sec4}

\subsection{\texorpdfstring{$f\sigma_8$ errors}{fσ8 errors}}

We computed the Fisher matrix of the parameters $f\sigma_8(z)$, $b_i\sigma_8(z)$, $\alpha_\parallel(z)$ and $\alpha_\perp(z)$ for redshifts $0.0<z<2.1$.  We then marginalised over the parameters to obtain the errors on $f\sigma_8$. The $f\sigma_8$ error bars can be seen for each redshift bin in Figure \ref{fig:fsigma8}. The strength of the constraint in each bin is determined by a combination of the associated number density and volume. It can be seen that the weakest constraint is given by the lowest redshift bin; this is due to the small volume sampled. High redshift bins also give poor constraints due to their lower number density.

\subsection{\texorpdfstring{Coupling constraints for a constant equation of state}{Coupling constraints for a constant equation of state}}
\subsubsection{\texorpdfstring{Fixed equation of state $w_0=-0.9$}{Fixed equation of state w0=-0.9}}

As described above, we obtained errors on the $\xi_i$ parameters using a Jacobian transform of the Fisher matrix. For the constant $w_0=-0.9$ model, we initially considered the errors on two parameters: $\xi_{\mathrm{low}}$, which is constant in time for the redshift range $0.0<z<2.1$, and $\xi_{\mathrm{high}}$. The projected constraints are shown in Table \ref{tab:iii}. In Figure \ref{fig:fsigma8}, we show how the evolution of $f\sigma_8(z)$ changes when computed with $\xi_{\mathrm{low}}$ and $\xi_{\mathrm{high}}$ values equal to the $3\sigma$ errors on the parameters. For this figure, we show only the positive error values to satisfy the condition $\xi \ge 0$. 

\begin{table}[tbp]
\centering
\begin{tabular}{|c||c|c||c|c|}
\hline
Redshift Range&$\xi$ Fisher Error&$\xi$ MCMC Error&$A$ Fisher Error&$A$ MCMC Error\\
\hline
$0.0<z<2.1$ & $0\pm 25.5$ & $\xi_{\mathrm{low}} < 11.4$ & $0\pm 2.55$ & $A_{\mathrm{low}} < 1.15$\\
$2.1<z<10$ & $0\pm 34.2$ & $\xi_{\mathrm{high}} < 15.3$ & $0\pm 3.42$ & $A_{\mathrm{high}} < 1.52$\\
\hline
$0.0<z<0.4$ & $0\pm 149$ & $\xi_1 < 107$ & $0\pm 14.9$ & $A_1 < 10.9$\\
$0.4<z<1.1$ & $0\pm 48.3$ & $\xi_2 < 21.0$ & $0\pm 4.83$ & $A_2 < 2.12$\\
$1.1<z<1.6$ & $0\pm 90.6$ & $\xi_3 < 23.1$ & $0\pm 9.06$ & $A_3 < 2.31$\\
$1.6<z<2.1$ & $0\pm 283$ & $\xi_4 < 29.1$ & $0\pm 28.3$ & $A_4 < 2.92$\\
$2.1<z<10$ & $0\pm 111$ & $\xi_{\mathrm{high}} < 11.6$ & $0\pm 11.1$ & $A_{\mathrm{high}} < 1.14$\\
\hline
\end{tabular}
\caption{\label{tab:iii} The calculated $1\sigma$ errors, in units of b/GeV, on each interaction strength parameter when we fix $w_0 = - 0.9$. We show both the expected errors from our Fisher forecasts as well as the $1\sigma$ upper bounds computed using MCMC with a $\xi \ge 0$ or $A \ge 0$ prior.}
\end{table}

\begin{figure}[tbp]
     \centering 
     \includegraphics[width=1 \textwidth]{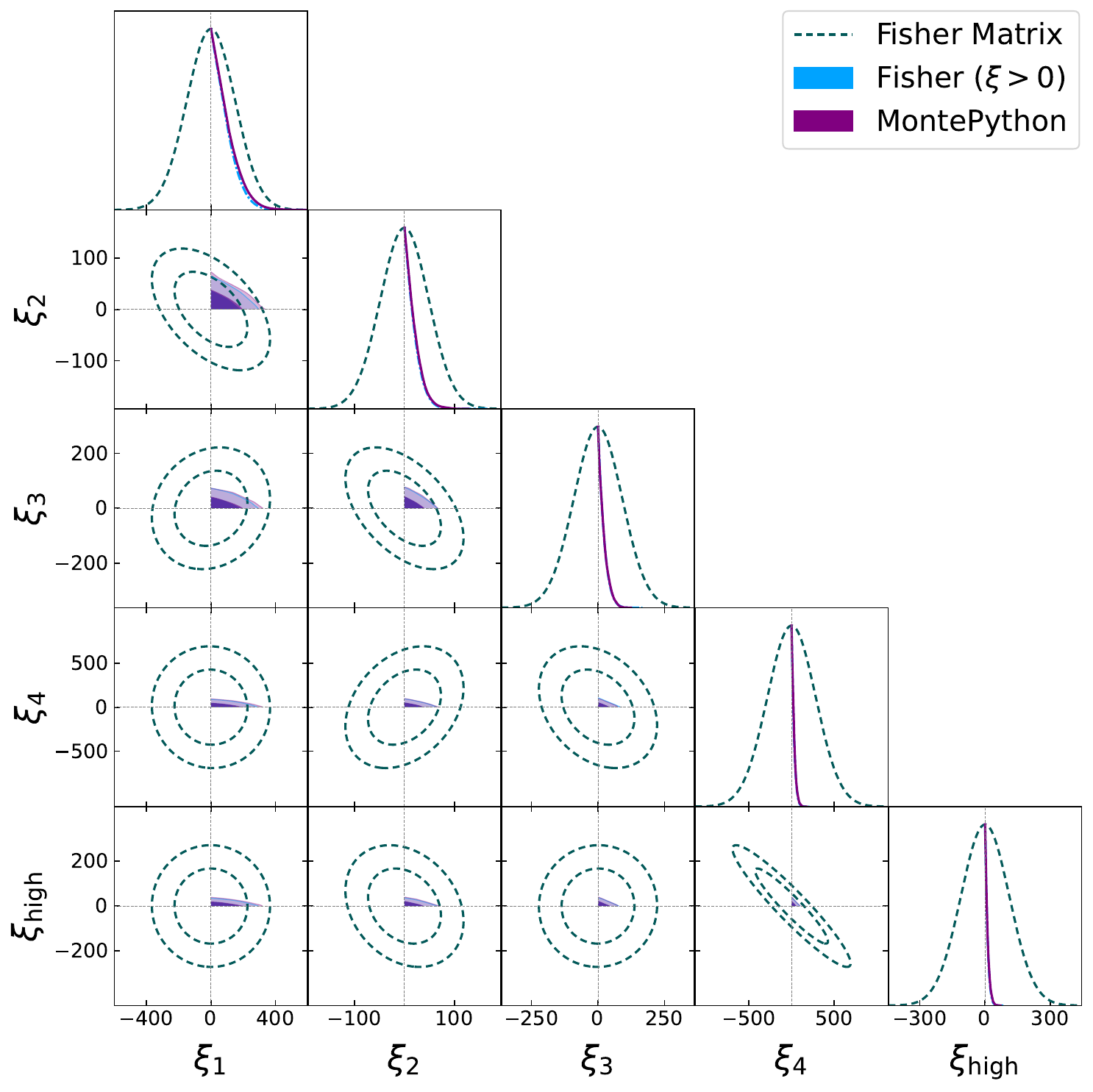}
     \hfill
     \caption{\label{fig:xi_w0_full} Forecasted $1\sigma$ and $2\sigma$ contours of the $\xi_i$ parameters, when modelled with a constant $w_0=-0.9$. We include the marginalised Fisher matrix results (dashed) as well as the constrained Fisher (blue) and \texttt{MontePython} (purple) results when computed with a $\xi \ge 0$ prior. The constrained Fisher and the MontePython results are very similar. }
\end{figure}

\begin{figure}[tbp]
     \centering 
     \includegraphics[width=0.7\textwidth]{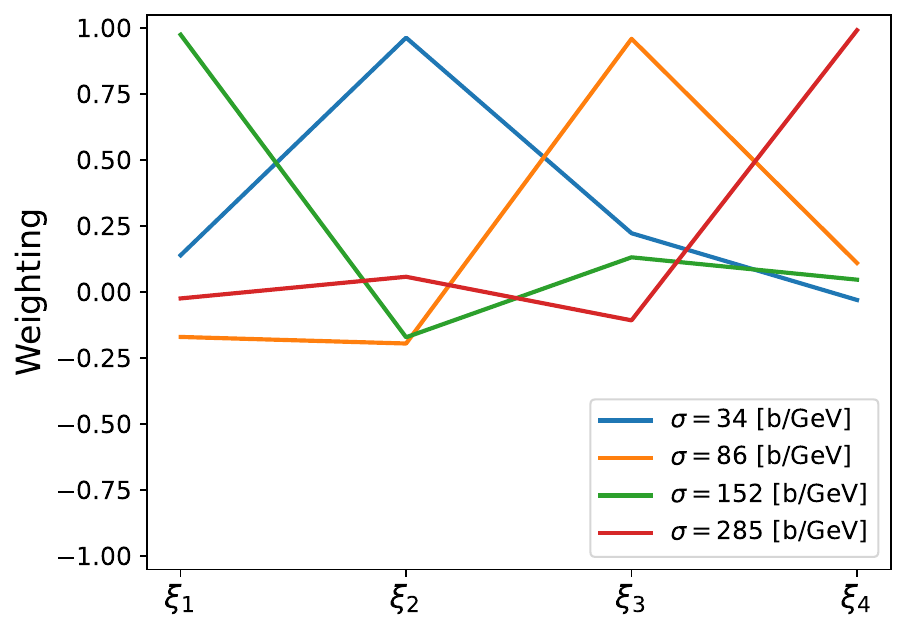}
     \hfill
     \caption{\label{fig:4} A PCA, where each mode is a linear combination of different parameters with different weighting. Here we show the eigenfunctions when $w_0=-0.9$ and is constant. }
\end{figure}

Following this, we investigated the errors for five $\xi_i$ parameters. 
In Figure \ref{fig:fsigma8}, we show the impact of changing these parameters individually on the evolution of $f\sigma_8(z)$.
These results can also be seen in Table \ref{tab:iii}. The Fisher matrix correlations between each pair of parameters from the resulting covariance matrix are shown by the dashed contours in Figure \ref{fig:xi_w0_full}. We make use of the Python package \texttt{GetDist} to plot these. In addition, we performed a Principal Component Analysis (PCA) on the $\xi_i$ covariance matrix to examine the relationships between the parameters. We focus on parameters $\xi_{1-4}$, as these span the redshift range that we expect to be directly constrained by data; we marginalise over $\xi_{\mathrm{high}}$ because, in the absence of high-redshift data, this parameter is degenerate with the amplitude of the primordial power spectrum, $A_s$. As a result, the constraint on $\xi_{\mathrm{high}}$ actually reflects the combined constraint on both parameters. Since we fix the value of $A_s$, our analysis provides tighter constraints on $\xi_{\mathrm{high}}$ than if $A_s$ were treated as a free parameter. Figure \ref{fig:4} shows the eigenvalues and eigenvectors of the covariance matrix, as well as how they relate to each parameter. It can be seen that the best constrained eigenfunction is dominated by $\xi_2$, has similar contributions from $\xi_1$ and $\xi_3$ and has the lowest contribution from $\xi_4$. The second and third best constrained eigenfunctions are primarily influenced by $\xi_3$ and $\xi_1$, respectively, whereas the worst constrained eigenfunction is most strongly associated with $\xi_4$.
This aligns with the constraints seen in Table \ref{tab:iii}.

The Fisher constraints include a range of unphysical negative $\xi$ values. In order to account for the required positive parameter values, we created synthetic data, with errors on $f\sigma_8$ obtained by the Fisher analysis, to perform MCMC with \texttt{MontePython} and a $\xi \ge 0$ prior. This analysis provided much tighter constraints than the Fisher estimate, as can be seen in Table \ref{tab:iii} and Figure \ref{fig:xi_w0_full}. As a check, we also used the Python package \texttt{emcee} to sample the Gaussian likelihood for $\xi_i$ with a $\xi \ge 0$ prior, using the computed $\xi_i$ covariance matrix from the Fisher matrix \cite{Foreman-Mackey_2013}. The result of this is shown alongside the \texttt{MontePython} MCMC contours in Figure \ref{fig:xi_w0_full}. These two approaches will agree if the posterior of $\xi_i$ is Gaussian. Although this is not necessarily the case, it can be seen that the two MCMC methods produce very comparable results.

The strength of each parameter constraint depends on a number of factors, including the number density, bias and effective volume of each affected tracer. For example, the low effective volume of the BGS tracers results in a weaker constraint on $\xi_1$; whereas $\xi_2$ has a stronger constraint, as it affects both the BGS and LRG tracers. $\xi_4$ and $\xi_{\mathrm{high}}$ are more weakly constrained, as they additionally affect a redshift range where the density of dark energy is negligible compared to that of CDM. In Figure \ref{fig:xi_w0_full}, it can be seen that $\xi_4$ and $\xi_{\mathrm{high}}$ are highly negatively correlated. Figure \ref{fig:fsigma8} shows that both parameters have a similar effect on the growth rate at lower redshifts, while only differing significantly for a few high redshift data points. This means that as the value of $\xi_{\mathrm{high}}$ increases, the value of $\xi_4$ must decrease to compensate.

We also examined the effect of changing the upper redshift limit on the $\xi_{\mathrm{high}}$ parameter constraint, with the maximum redshift being reduced to $z=5$ or increased to $z=20$. It was found that the upper redshift limit makes minimal difference to the final result. This is to be expected, as the largest impact of the coupling occurs when the dark energy density is comparable to that of dark matter; above a redshift of $z=5$, the relative dark energy density is negligible. This can be demonstrated by changing the upper redshift limit when constraining $\xi_{\mathrm{high}}$ with a $\xi \ge 0$ prior. For upper redshift values of $z=5$ and $z=20$, the $1\sigma$ upper bound on $\xi_{\mathrm{high}}$ is found to be $\xi_{\mathrm{high}} < 20.4$ and $\xi_{\mathrm{high}} < 13.8$ respectively. These are not substantially different from the value found using an upper redshift limit of $z=10$, given in Table \ref{tab:iii}.

\subsubsection{\texorpdfstring{Varying equation of state $w_0$}{Varying equation of state w0}}

So far we have focused on the case where the equation of state is fixed at $w_0 = -0.9$; here we consider also allowing the equation of state to vary.  There are strong degeneracies between the couplings and the equation of state, as the latter impacts the structure growth even in the absence of any coupling and it also modulates the dark matter damping when the coupling is present (Eq \ref{eq:4}).  For this reason, previous analyses have often focused on constraining the combination $A = \xi (1+w_0)$.   

\begin{figure}[tbp]
     \centering 
     \includegraphics[width=0.7
     \textwidth]{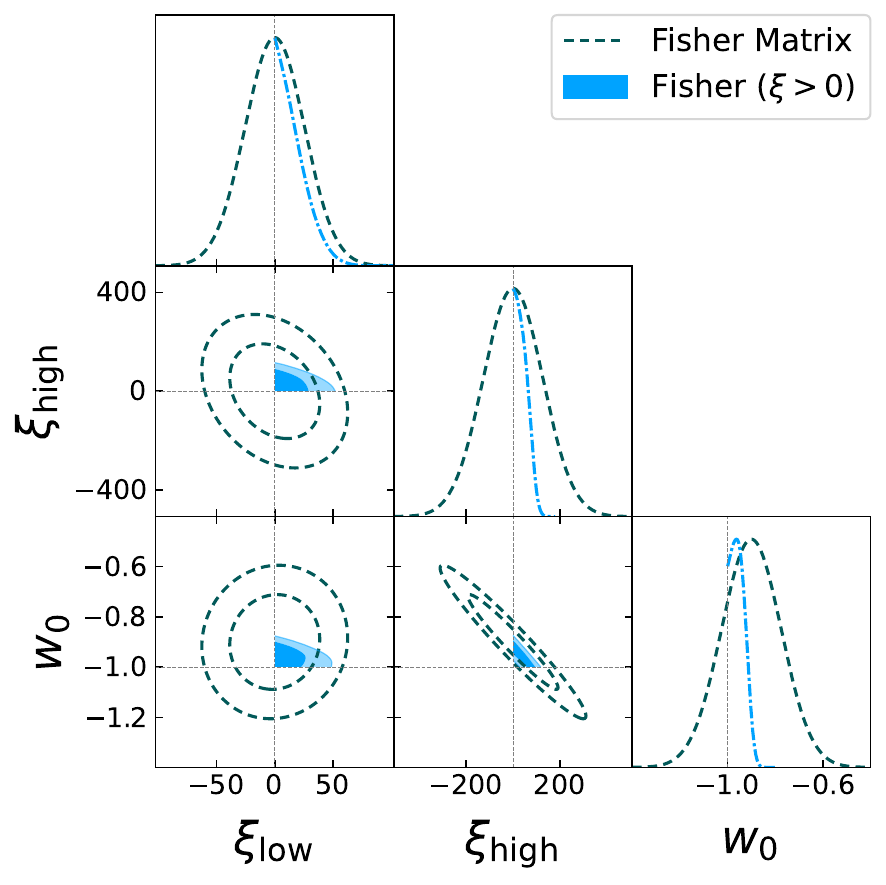}
     \hfill
     \caption{\label{fig:xi_w0} Forecasted constraints on the $\xi_{\mathrm{low}}$, $\xi_{\mathrm{high}}$ and $w_0$ parameters, where the fiducial model has $w_0=-0.9$. We include the marginalised Fisher matrix results as well as the Fisher results when constrained with $\xi \ge 0$ and $w_0>-0.999$ priors. We see that $w_0 = -1$ is allowed within $1\sigma$; in this limit, the $\xi$ parameterisation is not adequate, and it is better to constrain $A = \xi(1+w_0)$.}
\end{figure}

In Figure \ref{fig:xi_w0}, we show the Fisher forecast for $\xi_{\mathrm{low}}$, $\xi_{\mathrm{high}}$ and $w_0$. From this it can be seen that, due to a lack of high-redshift data, $\xi_{\mathrm{high}}$ and $w_0$ are highly correlated.  The equation of state is independently constrained by the AP constraints, but it can be seen that their current constraining power allows $w_0$ to approach a value of $-1$. This can lead to problems when trying to constrain the values of both $\xi$ and $w_0$, since $\xi$ can take much larger values when the $1+w$ factor in the coupling term approaches $0$ as $w_0$ approaches $-1$. For this reason, we now also examine the constraining power of future surveys on the interaction strength parameter $A(z)$, when split across the same number of redshift bins. This allows us to simultaneously forecast constraints on $w_0$ and the coupling strength of the interaction.  We note that when $w_0$ is fixed, the forecasted constraints on the $A_i$ parameters agree with the constraints on $\xi_i$ when multiplied by $1+w$, as can be seen in Table \ref{tab:iii}. 

We follow the same Fisher forecasting methodology that we used in the $\xi(z)$ case. When performing our MCMC analysis, we used a flat $A \ge 0$ prior. Given our prior of $w_0 > -1$, a negative value of $A$ would require a negative $\xi$, which is unphysical and would lead to instabilities in the perturbations. Additionally, we used a flat $w_0 > -0.999$ prior, chosen as the closest value to $w_0 = -1$ where the constraints remain consistent regardless of whether the common approximation is used that dark energy perturbations are negligible when $c_s = 1$. As expected, this parametrisation allows for constraints on both $A_i$ and $w_0$, while sharing the same correlations as the $\xi_i$ parameters. 

\begin{figure}[tbp]
     \centering 
     \includegraphics[width=1 \textwidth]{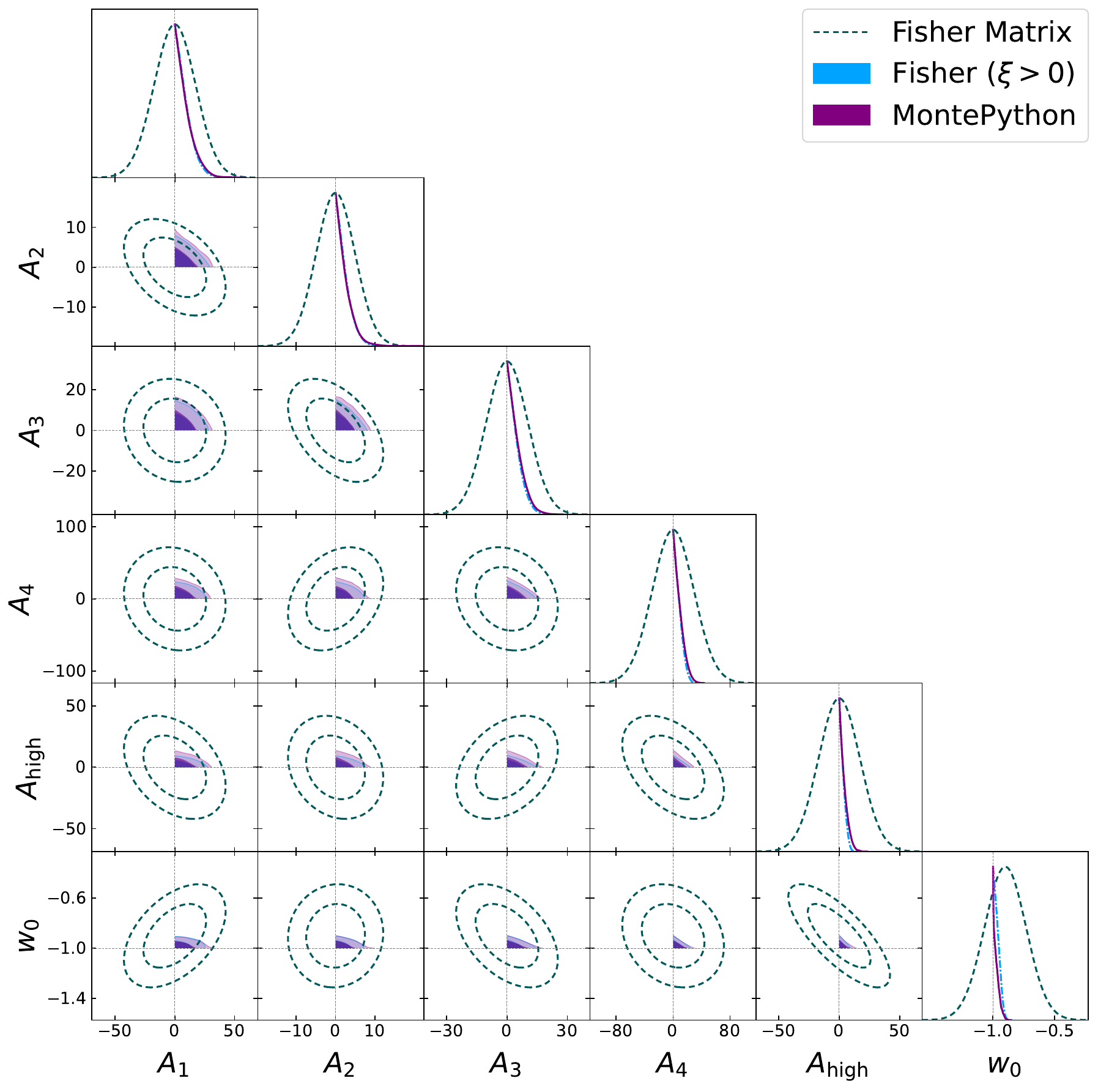}
     \hfill
     \caption{\label{fig:A_w0} Forecasted $1\sigma$ and $2\sigma$ contours of the $A_i$ and $w_0$=constant parameters, where the fiducial model has $w_0=-0.9$. We include the marginalised Fisher matrix results as well as the Fisher and \texttt{MontePython} results when computed with $A \ge 0$ and $w_0>-0.999$ priors.}
\end{figure}

Figure \ref{fig:A_w0} shows the Fisher forecast as well as the improvement in constraining power when using the chosen priors. The observed correlations between the $A_i$ parameters and $w_0$ arise from the growth rate suppression that results from greater values of $w_0$. As with the $\xi_i$ analysis, there is little difference between the MCMC results and the prior constrained Fisher forecasts. Our results show that the $A_i$ parameters are more useful than $\xi_i$ for extracting information about time-dependent momentum interactions from upcoming surveys, as they allow us to better distinguish the effects of the coupling and $w_0$ on the suppression of the growth rate at different redshifts.

We note that when the couplings are constrained such that $\xi_i \ge 0$, they can only fluctuate upwards from their fiducial values.  Since the couplings are largely anti-correlated with the equation of state, this results in biasing the inferred value of $w_0 \rightarrow -1$, as can be seen in Figures \ref{fig:xi_w0} and \ref{fig:A_w0}.  Such a bias could be significant when there is no clear detection of the coupling.

\subsection{\texorpdfstring{Coupling constraints for a thawing equation of state $w(z)$}{Coupling constraints for a thawing equation of state w(z)}}

\begin{figure}[tbp]
     \centering 
     \includegraphics[width=1 \textwidth]{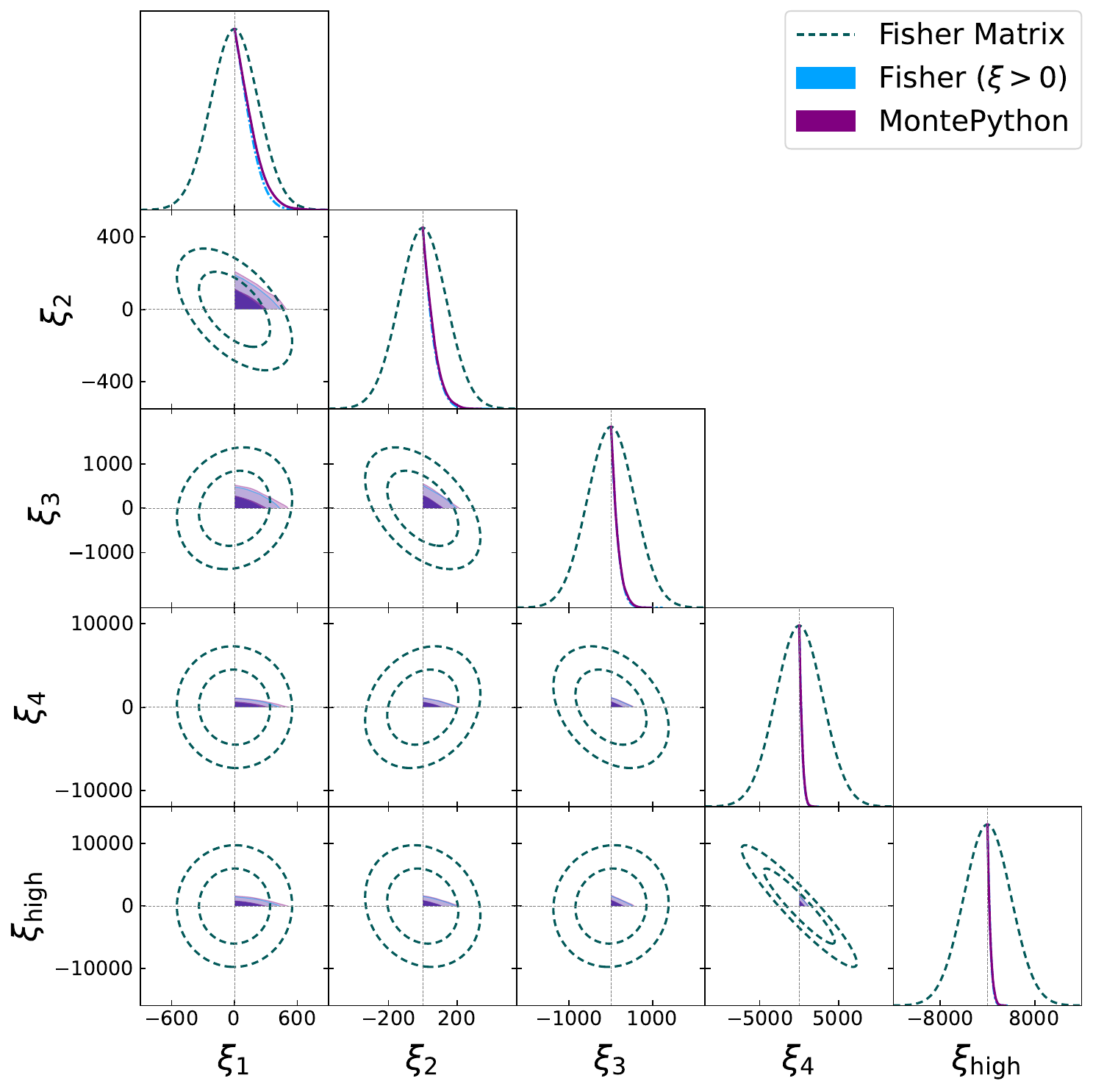}
     \hfill
     \caption{\label{fig:xi_kappa} Forecasted $1\sigma$ and $2\sigma$ contours of the $\xi_i$ parameters, when modelled with a fixed thawing equation of state, where $w(z= 0)= -0.9 $. We include the marginalised Fisher matrix results as well as the Fisher and \texttt{MontePython} results when computed with a $\xi \ge 0$ prior.}
\end{figure}

Following our analysis of the interaction in the case of a constant $w$, we also explored the coupling constraints for the thawing $w(z)$ model outlined above. The results of this can be seen in Table \ref{tab:iv}. For the thawing dark energy model we considered, the resulting constraints on $\xi_i$ are weaker due to the interaction's dependence on $\xi(1+w)$. In this model, $w(z)$ tends towards a value of $-1$ and $1+w \rightarrow 0$ at higher redshifts, which lessens the effect of the coupling. This has the greatest impact on $\xi_4$ and $\xi_{\mathrm{high}}$, where $1+w$ is closest to zero. As shown in Figure \ref{fig:xi_kappa}, the correlations between the parameters remain largely unchanged when compared to the constant $w$ case. Similar improvements to the parameter constraints are also observed when performing MCMC with a $\xi_i>0$ prior.

\begin{table}[tbp]
\centering
\begin{tabular}{|c||c|c||c|c|}
\hline
Redshift Range&$\xi$ Fisher Error&$\xi$ MCMC Error&$A$ Fisher Error&$A$ MCMC Error\\
\hline
$0.0<z<2.1$ & $0\pm 69.5$ & $\xi_{\mathrm{low}} < 49.5$ & $0\pm 2.53$ & $A_{\mathrm{low}} < 1.18$\\
$2.1<z<10$ & $0\pm 956$ & $\xi_{\mathrm{high}} < 731$ & $0\pm 4.19$ & $A_{\mathrm{high}} < 2.27$\\
\hline
$0.0<z<0.4$ & $0\pm 225$ & $\xi_1 < 164$ & $0\pm 14.8$ & $A_1 < 10.7$\\
$0.4<z<1.1$ & $0\pm 137$ & $\xi_2 < 61.7$ & $0\pm 4.83$ & $A_2 < 2.08$\\
$1.1<z<1.6$ & $0\pm 562$ & $\xi_3 < 160$ & $0\pm 9.47$ & $A_3 < 2.41$\\
$1.6<z<2.1$ & $0\pm 2978$ & $\xi_4 < 351$ & $0\pm 30.6$ & $A_4 < 3.35$\\
$2.1<z<10$ & $0\pm 3974$ & $\xi_{\mathrm{high}} < 502$ & $0\pm 14.9$ & $A_{\mathrm{high}} < 1.65$\\
\hline
\end{tabular}
\caption{\label{tab:iv} The calculated $1\sigma$ errors, in b/GeV, on each interaction strength parameter for a time-dependent thawing model, where we have fixed $w(z=0)=-0.9$. We show both the expected errors from our Fisher forecasts as well as the $1\sigma$ upper bounds computed using MCMC and a $\xi \ge 0$ or $A \ge 0$ prior.}
\end{table}

\begin{figure}[tbp]
     \centering 
     \includegraphics[width=0.7\textwidth]{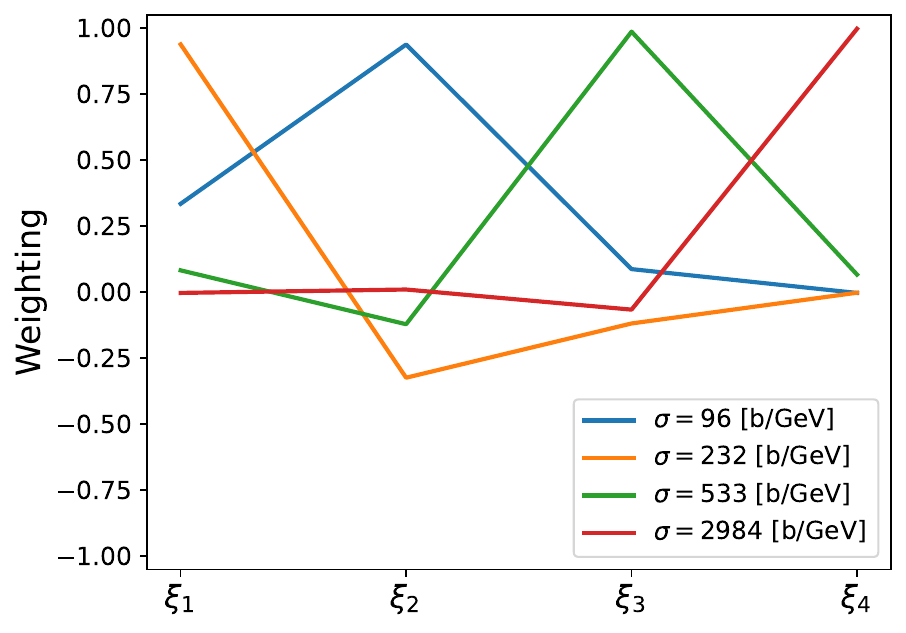}
     \hfill
     \caption{\label{fig:kappa_pca} Here we show the principal components for the thawing parameterisation, where $w \simeq -1$ at high redshifts and evolves with time to $w =-0.9$ at $z = 0$. The best constrained modes tend to shift to lower redshift bins due to $(1+w) \rightarrow 0 $ at higher redshifts. }
\end{figure}

The results of the PCA for this $\xi_i$ covariance matrix can be seen in Figure \ref{fig:kappa_pca}. When compared to Figure \ref{fig:4}, it can be seen that the eigenfunctions have weaker constraints. The best constrained eigenfunction is still dominated by $\xi_2$, whereas the second best constrained eigenfunction is now dominated by $\xi_1$. This matches the constraints seen in Table \ref{tab:iv} and is due to the weaker coupling strength at higher redshifts as $1+w\rightarrow0$.

\begin{figure}[tbp]
     \centering 
     \includegraphics[width=1 \textwidth]{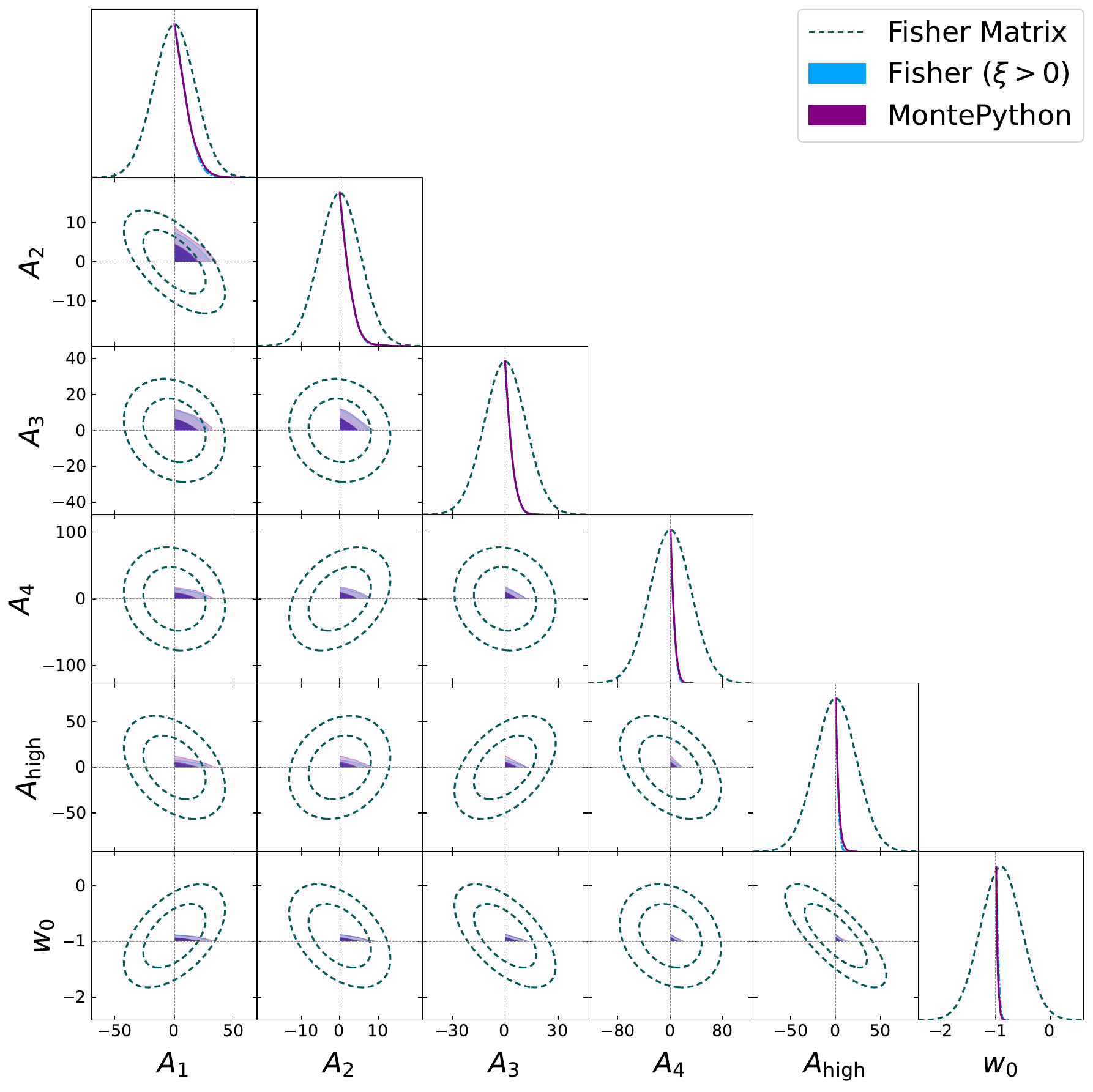}
     \hfill
     \caption{\label{fig:A_kappa} Forecasted $1\sigma$ and $2\sigma$ contours of the $A_i$ and $w_0 \equiv w(z=0)$ parameters, when modelled with a thawing $w(z)$. We include the marginalised Fisher matrix results as well as the Fisher and \texttt{MontePython} results when computed with $A \ge 0$ and $w_0>-0.99$ priors. }
\end{figure}

The parametrisation of $A(z)$ ensures that the $A_i$ parameters are just as well constrained by the data as in the constant $w$ case. This can be seen by comparing Tables \ref{tab:iii} and \ref{tab:iv}. There is a difference in how well the low redshift equation of state $w(z=0)$ is constrained by the data. For a constant equation of state, there is an observational impact at all redshifts. For thawing models however, their behaviour tends to converge at high redshifts regardless of the final equation of state, so that the only observational differences are at low redshifts. Thus, the constraint on $w(z=0)$ is weaker for thawing models than for constant equation of state models. As in the $\xi_i$ analysis, Figure \ref{fig:A_kappa} shows similar correlations to the constant $w_0$ case. The improvements in constraining power seen when performing MCMC with the chosen priors also remain consistent between the different $w$ models.

\section{Conclusions}
\label{sec5}

Here we have demonstrated the potential of the next generation experiments, such as a DESI-like survey, to constrain interactions between dark matter and dark energy. In particular, we have focused on models with pure momentum exchange, which can have the effect of suppressing structure growth; while such models do not affect the background expansion, their impact can be seen in measurements of redshift space distortions. By binning the coupling in redshift, we have also shown that the time-dependence of these interactions can be probed. 

The strength of the interactions generically depends on the combination $A=\xi(1+w(z))$, meaning that their impact can depend sensitively on the model of dark energy. For a constant equation of state, the coupling can be important at higher redshift. However, if, as is expected in thawing quintessence models, the dark energy approaches a cosmological constant behaviour ($w \rightarrow -1$) at high redshifts, then the impact of this coupling can be significantly decreased.  Such models also provide a challenge when attempting to constrain both $\xi(z)$ and $w_0$ at higher redshifts. Choosing instead to probe $A(z)$ provides a way of obtaining consistent constraints for $w_0$ and the interaction strength, regardless of dark energy model.  

We note in closing that some of the evidence for the $S_8$ tension has appeared to have weakened with recent analyses.  In particular, the most recent KiDS-Legacy constraints \cite{Wright:2025xka}, produced as this work was being finalised, shows significantly less tension with $\Lambda$CDM and the CMB data than their original analyses.  These questions should be greatly clarified by new experiments such as DESI, the Euclid satellite (\url{https://www.esa.int/Science_Exploration/Space_Science/Euclid}) and the Rubin-LSST survey (\url{https://rubinobservatory.org}).  Indeed, the DESI experiment is expected to release its second set of RSD results in the near future. Though these will not match the full forecast constraints assumed here, it will be interesting to combine them with other measurements, including weak lensing, Sunyaev-Zel'dovich (SZ) and other RSD measurements to constrain such momentum exchange models.


\acknowledgments

NC is supported by the UK Science and Technology Facilities Council (STFC) grant number ST/X508688/1 and funding from the University of Portsmouth. MB, RC and KK are supported by STFC grant
number ST/W001225/1. 

NC thanks Nathan Findlay and Ruiyang Zhao for helpful discussions when writing the Fisher forecasting code. We thank the anonymous referee for the helpful suggestions. Numerical computations were done on the Sciama High Performance Compute (HPC) cluster which is supported by the ICG, SEP-Net, and the University of Portsmouth. We also acknowledge the use of the Fisher forecast tool GoFish (\url{https://github.com/ladosamushia/GoFish}) as inspiration when writing our own Fisher forecasting code.

For the purpose of open access, we have applied a Creative
Commons Attribution (CC BY) licence to any Author Accepted
Manuscript version arising.







\bibliographystyle{JHEP}
\bibliography{references.bib}

\end{document}